\documentstyle[aps,epsfig]{revtex}
\begin{document}

\title{  Nuclear  absorption  and  $J/\psi$  suppression in Pb+Pb
collisions}

\author{\bf A. K. Chaudhuri\cite{byline}}
\address{Department of Physics,\\
Technion-Israel Institute of Technology,\\
Haifa 32000, Israel\\
and}
\address{ Variable Energy Cyclotron Centre\\
1/AF,Bidhan Nagar, Calcutta - 700 064\\}
\date{\today}

\maketitle

\begin{abstract}

We  have analyzed the NA58 data on $J/\psi$ suppression in Pb+Pb
collisions. $J/\psi$ production is assumed to be a two step
process, (i) formation of $c\bar{c}$ pair, which is accurately
calculable in QCD and (ii) formation of $J/\psi$ meson from   the
$c\bar{c}$ pair,   which   can   be   conveniently parameterized.
In  a pA/AA collision, a $c\bar{c}$ pair gain relative square
momentum as it passes through the nuclear medium and some of the
$c\bar{c}$ pairs can gain enough  momentum  to  cross the
threshold  to become  an  open  charm  meson,  leading  to
suppression in pA/AA collisions. A new prescription is proposed
for the gain in momentum square, consistent with Krammer process.
The model without any free parameter could explain the $E_T$
dependence of $J/\psi$ over Drell-Yan ratio.

\end{abstract}


In  relativistic  heavy  ion  collisions $J/\psi$ suppression has
been recognized as an important tool  to  identify  the  possible
phase transition to quark-gluon plasma. Because of the large mass
of  the  charm  quarks,  $c\bar{c}$ pairs are produced on a short
time scale. Their tight binding also makes them immune  to  final
state interactions. Their evolution probes the state of matter in
the  early  stage  of the collisions. Matsui and Satz \cite{ma86}
predicted that in presence of quark-gluon plasma  (QGP),  binding
of  $c\bar{c}$  pairs  into  a  $J/\psi$  meson will be hindered,
leading to the  so  called  $J/\psi$  suppression  in  heavy  ion
collisions  \cite{ma86}.  Over  the  years  several  groups  have
measured the $J/\psi$ yield in heavy ion collisions (for a review
of the data and the interpretations see Refs.  \cite{vo99,ge99}).
In  brief,  experimental  data  do show suppression, which could
be attributed to the conventional nuclear absorption, also present
in $pA$ collisions.

The latest data obtained by the NA50 collaboration \cite{na50} on
J/$\psi$ production in Pb+Pb collisions at 158 A GeV is the first
indication  of the anomalous mechanism of charmonium suppression,
which goes beyond  the  conventional  suppression  in  a  nuclear
environment.  The  ratio  of  J/$\psi$ yield to that of Drell-Yan
pairs decreases faster with $E_T$ in the most central  collisions
than  in  the  less  central ones. It has been suggested that the
resulting pattern can be understood in a  deconfinement  scenario
in  terms  of  successive  melting  of  charmonium  bound  states
\cite{na50}.  Essentially in a QGP like scenari, assuming all the
$J/\psi$ melts above a threshold density, NA50 data could be
explained as an effect of transverse energy fluctuations
\cite{bl00,ch01,ch02}. The data could be also be explained in the
comover approach without invoking QGP like scenario  \cite{ca00}.
Recently we have shown \cite{ch01b} that the NA50 data could be
well explained extending the model of Qiu et al \cite{qiu98}. In
this  model suppression is due to gain in relative square momentum
of $c\bar{c}$ pair, as it pass through the nuclear medium. Some of
the $c\bar{c}$ pair might gain enough momentum to cross the open
charm threshold.

In the present paper, we further analyze the problem and alter the
prescription used by Qiu et al \cite{qiu98} to calculate the gain
in the relative square momentum of the $c\bar{c}$ pair to be
consistent with theoretical understanding. We argue that square of
relative momentum gained by  $c\bar{c}$ pair for traversing a
length $L$ in nuclear medium goes as $L^2$ rather than $L$, as was
proposed by Qiu et al \cite{qiu98}.

We briefly describe the model of Qiu et al \cite{qiu98}.
Production of $J/\psi$ meson is assumed to be a two step process,
(i) production of $c\bar c$ pairs with relative momentum square
$q^2$, and (ii) formation  of $J/\psi$ mesons from the $c\bar{c}$
pairs. Step (i) can be accurately calculated  in  QCD
\cite{qiu98,be94}. The  second step, formation  of  $J/\psi$
mesons from initially compact $c\bar{c}$ pairs  is
non-perturbative. They  used a parametric form for the step (ii),
formation of $J/\psi$ from $c\bar{c}$ pairs. The $J/\psi$ cross
section in AB collisions, at centre of mass energy $\sqrt{s}$ was
then written as,

\begin{equation}\label{1} \sigma_{A+B \rightarrow J/\psi + X} (s)
= K \sum_{a,b} \int dq^2 \left( \frac{\hat \sigma_{ab \rightarrow
cc}}     {Q^2}     \right)    \int    dx_F    \phi_{a/A}(x_a,Q^2)
\phi_{b/B}(x_b,Q^2) \frac{x_a x_b}{x_a + x_b} \times  F_{c\bar{c}
\rightarrow J/\psi} (q^2), \end{equation}

\noindent  where  $\sum_{a,b}$  runs over all parton flavors, and
$Q^2 = q^2 +4 m_c^2$. The  $K$  factor  takes  into  account  the
higher  order corrections. The incoming parton momentum fractions
are fixed by kinematics and are $x_a
=(\sqrt{x^2_F+4Q^2/s}+x_F)/2$               and              $x_b
=(\sqrt{x^2_F+4Q^2/s}-x_F)/2$.  Subprocess cross sections can be found in
\cite{be94}

In  a nucleon-nucleus/nucleus-nucleus  collision,  the produced
$c\bar{c}$ pairs interact with nuclear medium before they  exit.
Observed  anomalous nuclear enhancement of the momentum imbalance
in dijet production led Qiu, Vary and Zhang \cite{qiu98} to argue
that  the  interaction  of  a  $c\bar{c}$ pair   with   nuclear
environment,  increases  the  square  of the  relative  momentum
between the $c\bar{c}$ pair. They prescribed that if the pair
traverses a length $L$ in nuclear medium, then the relative square
momentum $q^2$ in the transition probability $F_{c\bar{c}
\rightarrow J/\psi}(q^2)$ should be changed to,

\begin{equation} \label{3a}
q^2 \rightarrow q^2+ \varepsilon^2 L
\end{equation}

\noindent with $\varepsilon^2$ being the square of relative
momentum gained by the pair per unit length of nuclear medium.
However, if we treat momentum gain as a random walk, then for a
Krammer like process, one easily obtain,

\begin{equation} \label{3b}
<(p_f-p_i)^2> \propto <(x_f-x_i)^2>
\end{equation}

In other words, for a Krammer like process, square of momentum
gain goes as $L^2$ rather than $L$ as in Eq.\ref{3a}. Accordingly,
we suggest that for traversing a length $L$ in nuclear medium, the
relative momentum $q^2$ in the transition probability $F_{c\bar{c}
\rightarrow J/\psi}(q^2) $ should be changed to

\begin{equation}
q^2 \rightarrow q^2 + \varepsilon^2 L^2
\end{equation}

The parameter $\varepsilon^2$ was obtained by fitting NA50 data
\cite{ab97} on $J/\psi$ production in pA and AA collisions. For
the transition probability, we use the following form
\cite{qiu98,ch01b},

\begin{equation} \label{4} F_{c \bar{c} \rightarrow J/\psi} (q^2)
=  N_{J/\psi} \theta(q^2) \theta({4m^\prime}^2 - 4 m_c^2 -q^2) (1
-   \frac{q^2}{{4m^\prime}^2   -   4   m_c^2  })^{\alpha_F},
\end{equation}

\noindent and we have used CTEQ5M parton distribution function \cite{cteq5}
 in the
calculation. In Fig.1, we have compared NA50 experimental data
\cite{ab97} with the best fitted curve. Except for the pd
reaction, all the data points are correctly reproduced. Cross
section for  pd reaction is underpredicted. For a Krammer process,
Eq.\ref{3b} ($ \Delta p^2 \propto \Delta x^2$) is obtained with
the assumption of long interval of time i.e. for large interval of
length. For pd collisions, effective length is small for the
equation to be valid.

We now apply the model to obtain transverse energy dependence of
the $J/\psi$ over Drell-Yan ratio. The  Drell-Yan  pairs  do not
suffer final state interactions and the cross section at an impact
parameter ${\bf b }$ as a function of $E_T$ can be written as,

\begin{equation}    \label{5}    d^2\sigma^{DY}/dE_T    d^2b    =
\sigma_{NN}^{DY} \int d^2s  T_A({\bf  s})  T_B({\bf  s}-{\bf  b})
P(b,E_T), \end{equation}

\noindent where $\sigma_{NN}^{DY}$ is the Drell-Yan cross section
in  $NN$  collisions. All the nuclear information is contained in
the nuclear thickness function,  $T_{A,B}({\bf  s})  (=  \int  dz
\rho_{A,B}({\bf  s},z)$.  Presently  we  have  used the following
parametric form for $\rho_A(r)$ \cite{bl00},

\begin{equation}                                        \label{6}
\rho_A(r)=\frac{\rho_0}{1+exp(\frac{r-r_0}{a})} \end{equation}

\noindent with $a=0.53 fm$, $r_0=1.1A^{1/3}$. The central density
is  obtained  from  $\int  \rho_A(r)  d^3r  =  A$. In Eq.\ref{5},
$P(b,E_T)$ is the  probability  to  obtain  $E_T$  at  an  impact
parameter  $b$.  Geometric  model  has  been  quite successful in
explaining  the  transverse  energy  as  well   as   multiplicity
distributions  in  $AA$  collisions  \cite{ch90,ch93}. Transverse
energy distribution in Pb+Pb collisions also could  be  described
in  this model \cite{ch01}. In this model, $E_T$ distribution is
written in terms of $E_T$ distribution in NN collisions. One also
assume that the Gamma distribution, with parameters $\alpha$  and
$\beta$  describe the $E_T$ distributions in NN collisions. Pb+Pb
data on $E_T$ distribution could be fitted with $\alpha =3.46 \pm
0.19$ and $\beta = 0.379 \pm 0.021$ \cite{ch01}.

While  Drell-Yan  pairs  do  not suffer interactions with nuclear
matter, the $J/\psi$ mesons do. In the model  suppression factor
depend  on  the length traversed  by  the $c\bar{c}$ mesons   in
nuclear   medium. Consequently,  we write  the $J/\psi$ cross
section at an impact parameter ${\bf b}$ as,

\begin{equation}   \label{7}   d^2\sigma^{J/\psi}/dE_T   d^2b   =
\sigma_{NN}^{J/\psi}  \int  d^2s  T_A({\bf  s})  T_B({\bf   s-b})
S(L({\bf b,s})) P(b,E_T), \end{equation}

\noindent  where  $\sigma_{NN}^{J/\psi}$  is  the  $J/\psi$ cross
section  in  $NN$  collisions  and  $S(L({\bf  b,s}))$   is   the
suppression  factor  due to passage through a length L in nuclear
environment. At an impact parameter ${\bf b}$ and at point  ${\bf
s}$, the transverse density can be calculated as,

\begin{equation}  \label{8}  n({\bf  b,s})  =  T_A({\bf  s}) [1 -
e^{-\sigma_{NN}  T_B({\bf  b-s})}]  +   T_B({\bf   b-s})   [1   -
e^{-\sigma_{NN} T_A({\bf s})}], \end{equation}

\noindent  and  the length $L({\bf b,s})$ that the $J/\psi$ meson
will traverse can be obtained as,

\begin{equation}  \label{9}  L({\bf  b,s})=n({\bf  b,s})/2 \rho_0
\end{equation}

Suppression  factor  $S(L({\bf  b,s})$  is then easily calculated
using Eq.\ref{1}.

Fluctuations  of  $E_T$,  at  a  fixed  impact parameter plays an
important role in the explanation of NA50 data. Following Blaizot
et  al \cite{bl00}, we take into account   $E_T$  fluctuations by
the following replacement,

\begin{equation}
L({\bf b,s}) \rightarrow L({\bf b,s}) E_T/E_T(b)
\end{equation}

In  Fig.2,  we  have  compared the model prediction with the NA50
experimental data on $J/\psi$ over DY ratio. The  solid  line  is
the  ratio  obtained  with  out  any $E_T$ fluctuations. The data
are   reproduced upto the knee of the $E_T$ distribution. Beyond
the knee of the $E_T$ distribution, model prediction saturates,
while the data shows rapid fall of the ratio. The dashes line is
the model prediction obtained including the $E_T$ fluctuations.
The model without any {\em free parameter} give excellent
description of the data, throughout the $E_T$ range. The
results clearly shows that the rapid fall of the $J/\psi$ over
Drell-Yan  ratio beyond the knee of the $E_T$ distribution is due
to $E_T$ fluctuation only. Present calculation also establishes that
it is not essential to assumed a deconfined scenario to
explain the NA50 data on $J/\psi$ to Drell-Yan ratio. Nuclear
absorption alone is capable of explaining it.

To summarize, we  have analyzed the NA50 data on transverse energy
distribution of $J/\psi$ in  Pb+Pb  collisions  at  CERN SPS.  In
the model, $c\bar{c}$  pairs  gain  relative  square  momentum
$\varepsilon^2$ as it travel a square of length $L^2$ through the
nuclear environment. Suppression occurs as some of the pairs can
gain enough momentum  to cross the threshold to become an open
charm meson. The parameters of  the model  were fixed by fitting
experimental $J/\psi$  cross section in pA and AA collisions. The
model could  very  well describe the transverse   energy
dependence of $J/\psi$ over Drell-Yan ratio upto the knee of the
$E_T$ distribution. It explains the data, throughout the $E_T$
range if $E_T$ fluctuations are included.

The author would like to thank  the Lady Davis Fellowship trust
for supporting his visit to Technion.

\begin{figure}[h]
\centerline{\psfig{figure=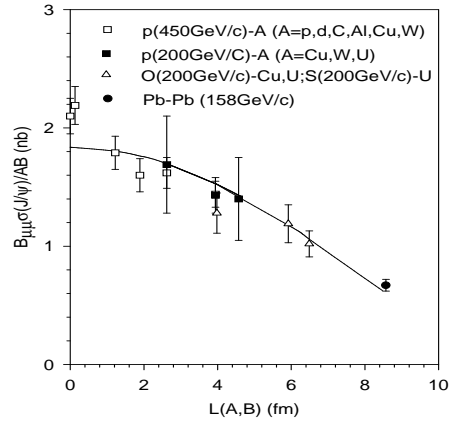,height=8cm,width=7cm}}
\vspace{-1cm}  \caption{
Total  $J/\psi$  cross  sections with the
branching ratio to $\mu^+\mu^-$ in proton-nucleus, proton-nucleus
and nucleus-nucleus collisions, as a function  of  the  effective
nuclear length L(A,B).} \end{figure}

\begin{figure}[h]
\centerline{\psfig{figure=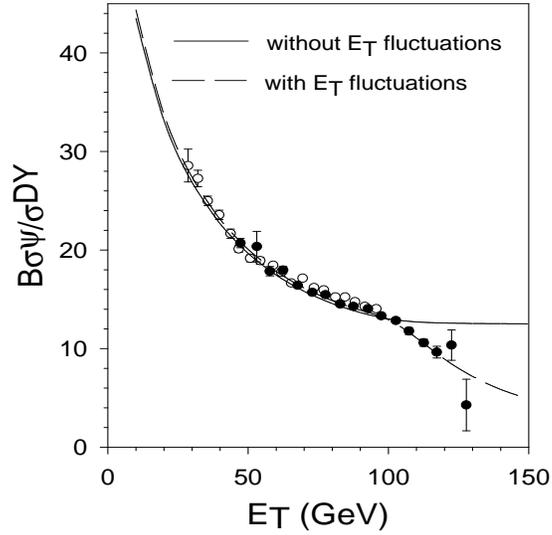,height=10cm,width=8cm}}
\vspace{-1cm}  \caption{Open  and closed circles are the $J/\psi$
to  Drell-Yan  ratio  in  a  Pb+Pb  collision  obtained  by  NA50
collaboration  in  1996 and 1998 respectively. The solid line is
the model prediction obtained without any $E_T$ fluctuations and
the dashed line is obtained including $E_T$ fluctuations.}
\end{figure}


\begin{references}  \bibitem[*]{byline}e-mail:akc@veccal.ernet.in
\bibitem{ma86} T. Matsui and H. Satz, Phys. Lett. B178,416(1986).
\bibitem{vo99}  R.  Vogt,  Phys.   Reports,   310,   197   (1999)
\bibitem{ge99}C.  Gerschel  and  J. Hufner, Ann. Rev. Nucl. Part.
Sci. 49, 255  (1999).
\bibitem{na50}NA50  collaboration,  M.  C.
Abreu  {\em  et al.} Phys. Lett. B 477,28(2000)
\bibitem{bl00} J.
P. Blaizot, P. M. Dinh and  J.Y.  Ollitrault,  Phys.  Rev.  Lett.
85,4012(2000).

\bibitem{ch01} A. K. Chaudhuri, hep-ph/0102038 ,
Phys. Rev. C64,054903 (2001).

\bibitem{ch02}A. K. Chaudhuri, Phys. Lett. B527,80(2002).

\bibitem{ca00}A. Capella, E. G.  Ferreiro
and  A.  B.  Kaidalov,  hep-ph/0002300,  Phys. Rev. Lett. 85,2080
(2000).


\bibitem{ch01b} A. K. Chaudhuri, hep-ph/0109141.
\bibitem{qiu98}  J.  Qiu, J. P. Vary, X. Zhang,
hep-ph/9809442, Nucl. Phys. A698, 571 (2002).
\bibitem{be94} C. J. Benesh, J.  Qiu  and  J.  P.
Vary Phys.Rev.C50:1015,(1994).


\bibitem{cteq5} CTEQ Collaboration
(H.L.   Lai   {\em   et   al.}).   Eur.Phys.J.C12,  375,  (2000).
\bibitem{ab97} M. C. Abreu {\em et al.}, Phys.  Lett.  B410,  337
(1997).

 \bibitem{ch90}  A.  K.  Chaudhuri,  Nucl.  Phys. A515,736(1990).
\bibitem{ch93}  A.  K.  Chaudhuri,  Phys.  Rev.   C47,2875(1993).



\end{references}
\end{document}